\documentstyle[osa,manuscript]{revtex}

\begin{document}
\title{Elasticity of Thin Rods with Spontaneous Curvature
and Torsion---Beyond Geometrical Lines }
\author{Aleksey D. Drozdov$^{1}$ and Yitzhak Rabin$^{2}$}
\address{$^{1}$ Institute for Industrial Mathematics, 4 Hanachtom Street\\
Beersheba, 84311 Israel\\
$^{2}$ Department of Physics, Bar-Ilan University\\
Ramat-Gan, 52900 Israel}
\maketitle

\begin{abstract}
We study three-dimensional deformations of thin inextensible elastic rods
with non-vanishing spontaneous curvature and torsion. In addition to the
usual description in terms of curvature and torsion which considers only the
configuration of the centerline of the rod, we allow deformations that
involve the rotation of the rod's cross-section around its centerline. We
derive new expressions for the mechanical energy and for the force and
moment balance conditions for the equilibrium of a rod under the action of
arbitrary external loads. Several illustrative examples are studied and the
connection between our results and recent experiments on the stretching of
supercoiled DNA molecules is discussed.
\end{abstract}

\pacs{87.15.La, 46.70.Hg}

\section{Introduction}

Recent experimental advances in the art of manipulation of single DNA
molecules and of rigid protein assemblies such as actin filaments, etc.,
have led to an outbreak of theoretical activity connected with the
elasticity of thin rods$^{1-19}$. One of the most intriguing theoretical
questions related to the deformation of DNA concerns the coupling between
bending and twist in the mechanical energy of the polymer. The problem is
usually considered in the following terms: at the first step, the thin rod
which models the molecule is replaced by its centerline. With each point of
the line (specified by its position along the contour $\xi $) one associates
a triad of unit vectors: the tangent to the line (${\bf t}$), the principal
normal (${\bf n}$) which lies in the plane defined by the tangents at points 
$\xi $ and $\xi +d\xi ,$ and the binormal (${\bf b}$) which is orthogonal to
both ${\bf t}$ and ${\bf n}$. As one moves along the line, the triad rotates
and this rotation is described by the Frenet--Serret equations in which the
``rate'' of rotation of each unit vector is determined by two parameters:
the local curvature $\kappa $ and the local torsion $\omega $ (sometimes
referred to as writhe)\cite{SH94}. In order to relate this purely
geometrical picture to the elastic response of real rods, one has to specify
the physical properties of the rod in a stress--free (undeformed) reference
state and to write the energy as a quadratic expansion in deviations from
this state. In the classical theories of thin elastic rods\cite{LL80} one
usually assumes that the reference state corresponds to a straight untwisted
rod (with vanishing spontaneous curvature $\kappa _{0}$ and spontaneous
torsion $\omega _{0}$) and the mechanical energy density is written as a sum
of terms proportional to $\kappa ^{2}$ and $\omega ^{2}$. The generalization
to the case of non--vanishing spontaneous curvature and torsion is then done
by requiring that the strain energy density per unit length $U$ is minimized
for $\kappa =\kappa _{0}$ and $\omega =\omega _{0}$ which leads to the
expression 
\begin{equation}
U=\frac{1}{2}\biggl [A_{1}(\kappa -\kappa _{0})^{2}+A_{2}(\omega -\omega
_{0})^{2}\biggr ].
\end{equation}
Here $A_{1}$ and $A_{2}$ are material parameters (products of elastic moduli
and moments of inertia).

Although Eq. (1) has been employed in a number of studies\cite
{HH91,GL92,MKJ95,MS95,GT97,GPW98,LGK98,HZ99}, its validity has been
questioned by several authors \cite{KLN97,Mar97,Mar98,MN98}, who argued that
it fails to describe, even qualitatively, the experimental data on
torsionally constrained DNA \cite{MN98,SAB96}. To account for the coupling
between bending and twist observed in experiment, extra terms are
conventionally added to the mechanical energy density,
Eq. (1), {\it by hand}.

The objective of this work is to derive an expression for the mechanical
energy and obtain the equations which determine the mechanical equilibrium
of a rod subjected to arbitrary forces and moments. This is done using a new
form of the displacement field, which accounts for both the deformation of
the centerline and the rotation of the cross-section around this line (i.e.,
twist). Instead of using ad hoc assumptions about the form of the coupling
between bending and twist, we will use standard methods of the theory of
elasticity in order to derive the correct form of the coupling.

In this work we will consider cylindrical rods with circular
cross--sections. Although, at first sight, this case appears to be simpler
than that of rods with asymmetric cross--sections, the reverse is true:
while in the asymmetric case one can introduce a triad of vectors associated
with the principal axes of inertia, which can rotate at a different rate
than the Frenet triad, no such natural choice is possible in the symmetric
case which therefore requires a more careful analysis.

The exposition is organized as follows. Section 2 deals with geometry of
deformation. The strain energy density of a rod is introduced in Section 3.
Stress--strain relations are developed in Section 4. In Section 5, force and
moment balance equations which describe the mechanical equilibrium of thin
rods are derived. Several examples which illustrate the different aspects of
the interaction between elongation, torsion and twist, are discussed in
Section 6. Finally, in Section 7 we discuss the connection between our
results and other theoretical and experimental works and outline directions
for future research.

\section{Geometry of deformation}

A long chain is modeled as an elastic rod with length $L$ and a circular
cross-section ${\cal S}$ with radius $a\ll L$. Denote by $\xi $ the
arc--length of the centerline of the rod in the reference (stress-free)
configuration. Let ${\bf R}_{0}(\xi )$ be the radius vector of the
longitudinal axis and ${\bf t}_{0}(\xi )=d{\bf R}_{0}/d\xi $ the unit
tangent vector in the reference state. The unit normal vector ${\bf n}%
_{0}(\xi )$ and the unit binormal vector ${\bf b}_{0}(\xi )$ are introduced
by the conventional way. These vectors obey the Frenet--Serret equations
with given spontaneous curvature $\kappa _{0}(\xi )$ and torsion $\omega
_{0}(\xi )$: 
\begin{equation}
\frac{d{\bf t}_{0}}{d\xi }=\kappa _{0}{\bf n}_{0},\qquad \frac{d{\bf n}_{0}}{%
d\xi }=\omega _{0}{\bf b}_{0}-\kappa _{0}{\bf t}_{0},\qquad \frac{d{\bf b}%
_{0}}{d\xi }=-\omega _{0}{\bf n}_{0}.
\end{equation}
Points of the rod refer to Lagrangian coordinates $\{\xi _{i}\}$, where $\xi
_{1},\xi _{2}$ are Cartesian coordinates in the cross-sectional plane with
unit vectors ${\bf n}_{0}$ and ${\bf b}_{0}$ and $\xi _{3}=\xi $, 
\begin{equation}
{\bf r}_{0}(\xi _{1},\xi _{2},\xi )={\bf R}_{0}(\xi )+\xi _{1}{\bf n}%
_{0}(\xi )+\xi _{2}{\bf b}_{0}(\xi ).
\end{equation}
It follows from Eqs. (2) and (3) that the covariant base vectors in the
reference configuration, ${\bf g}_{0k}=\partial {\bf r}_{0}/\partial \xi
_{k} $ are given by 
\begin{equation}
{\bf g}_{01}={\bf n}_{0},\qquad {\bf g}_{02}={\bf b}_{0},\qquad {\bf g}%
_{03}=(1-\kappa _{0}\xi _{1}){\bf t}_{0}+\omega _{0}(\xi _{1}{\bf b}_{0}-\xi
_{2}{\bf n}_{0}).
\end{equation}
The position of the longitudinal axis of the rod in the actual (deformed)
configuration is determined by the radius vector ${\bf R}(\xi )$. Following
the conventional theories of rods, see, e.g., \cite{BZ79}, we assume that
the longitudinal axis is inextensible, which means that $\xi $ remains the
arc--length in the actual configuration (for attempts to account for the
extensibility of the longitudinal axis, see \cite{Odi95,KLN97,Mar97,Mar98}).
The unit tangent vector in the actual configuration ${\bf t}=d{\bf R}/d\xi $
together with the unit normal vector ${\bf n}$ and the unit binormal vector $%
{\bf b}$ satisfy the Frenet--Serret equations 
\begin{equation}
\frac{d{\bf t}}{d\xi }=\kappa {\bf n},\qquad \frac{d{\bf n}}{d\xi }=\omega 
{\bf b}-\kappa {\bf t},\qquad \frac{d{\bf b}}{d\xi }=-\omega {\bf n}.
\end{equation}
For Kirchhoff rods \cite{GT97}, the radius vector of an arbitrary point is
represented as an expansion in the coordinates $\xi _{1}$ and $\xi _{2}$: 
\begin{equation}
{\bf r}(\xi _{1},\xi _{2},\xi )={\bf R}(\xi )+\xi _{1}{\bf n}(\xi )+\xi _{2}%
{\bf b}(\xi ).
\end{equation}
The functional form of Eq. (6) implies that any cross-section remains planar
and perpendicular to the centerline of the rod, even in the actual deformed
configuration. Furthermore, it also implies that any cross--section rotates
rigidly with the longitudinal axis and therefore Eq. (6) does not allow for
the possibility of a twist of the cross--section with respect to the
centerline of the rod. Since the latter assumption has no physical basis 
\cite{GL92}, we relax it by introducing a more general displacement field 
\begin{equation}
{\bf r}(\xi _{1},\xi _{2},\xi )={\bf R}(\xi )+(\xi _{1}\cos \alpha -\xi
_{2}\sin \alpha ){\bf n}(\xi )+(\xi _{1}\sin \alpha +\xi _{2}\cos \alpha )%
{\bf b}(\xi ),
\end{equation}
where $\alpha (\xi )$ is the rotation angle around the centerline of the
rod. From here on we will refer to this rotation as ``twist'' and will
reserve the terms ``torsion'' and ``writhe'' to describe the
three-dimensional geometry of bending of the centerline of the rod.

The covariant base vectors ${\bf g}_{k}=\partial {\bf r}/\partial \xi _{k}$
are given by 
\begin{eqnarray}
{\bf g}_{1} &=&\cos \alpha {\bf n}+\sin \alpha {\bf b}, \qquad {\bf g}%
_{2}=-\sin \alpha {\bf n}+\cos \alpha {\bf b},  \nonumber \\
{\bf g}_{3} &=& \Bigl [ 1 -\kappa (\xi _{1}\cos \alpha -\xi _{2}\sin \alpha
) \Bigr]{\bf t}  \nonumber \\
&&+\Bigl(\omega +\frac{d\alpha }{d\xi }\Bigr) \Bigl [ -(\xi _{1}\sin \alpha
+\xi _{2}\cos \alpha ){\bf n} +(\xi _{1}\cos \alpha -\xi _{2}\sin \alpha )%
{\bf b}\Bigr].
\end{eqnarray}
The contravariant base vectors ${\bf g}^{k}$ are found from Eq. (8) and the
equality ${\bf g}_{i}\cdot {\bf g}^{j}=\delta _{i}^{j}$, where the dot
stands for inner product and $\delta _{i}^{j}$ is the Kronecker delta.
Simple calculations result in 
\begin{equation}
{\bf g}^{1}=\cos \alpha {\bf n}+\sin \alpha {\bf b}+C_{1}{\bf t}, \qquad 
{\bf g}^{2}=-\sin \alpha {\bf n}+\cos \alpha {\bf b}+C_{2}{\bf t}, \qquad 
{\bf g}^{3}=C_{3}{\bf t},
\end{equation}
where 
\begin{eqnarray}
&& A_{n}=-\Bigl(\frac{d\alpha }{d\xi } +\omega \Bigr)(\xi _{1}\sin \alpha
+\xi _{2}\cos \alpha ), \qquad A_{b}=\Bigl(\frac{d\alpha }{d\xi }+\omega
\Bigr ) (\xi _{1}\cos \alpha -\xi _{2}\sin \alpha ),  \nonumber \\
&& A_{t}=1-\kappa (\xi _{1}\cos \alpha -\xi _{2}\sin \alpha ), \qquad
C_{1}=- \frac{1}{A_{t}}(A_{n}\cos \alpha +A_{b}\sin \alpha ),  \nonumber \\
&& C_{2}=\frac{1}{A_{t}}(A_{n}\sin \alpha -A_{b}\cos \alpha ), \qquad C_{3}= 
\frac{1}{A_{t}}.
\end{eqnarray}
One can proceed to calculate the energy of deformation using the
displacement gradient either in the deformed, $\mbox{$\boldmath \nabla$}{\bf %
r}_{0}$, or in the reference, $\mbox{$\boldmath \nabla$}_{0}{\bf r}$, state.
Both approaches result in the same expression for the mechanical energy. We
will use the displacement gradient in the actual configuration $%
\mbox{$\boldmath \nabla$}{\bf r}_{0}$, because the corresponding strain
tensor is connected with the stress tensor (always defined with respect to
the coordinates in the deformed state) by conventional constitutive
equations. It follows from Eqs. (4) and (9) that the tensor $%
\mbox{$\boldmath \nabla$}{\bf r}_{0}={\bf g}^{k}{\bf g}_{0k}$ is given by 
\begin{eqnarray}
\mbox{$\boldmath \nabla$}{\bf r}_{0} &=&\cos \alpha ({\bf n}{\bf n}_{0}+{\bf %
b}{\bf b}_{0}) +\sin \alpha ({\bf b}{\bf n}_{0}-{\bf n}{\bf b}_{0}) 
\nonumber \\
&& +(C_{1}-C_{3}\omega _{0}\xi _{2}){\bf t}{\bf n}_{0} +(C_{2}+C_{3}\omega
_{0}\xi _{1}){\bf t}{\bf b}_{0} +C_{3}(1-\kappa _{0}\xi _{1}){\bf t}{\bf t}%
_{0}.
\end{eqnarray}
As a measure of deformation, the Almansi tensor \cite{Dro96} ${\bf A}=%
\mbox{$\boldmath \nabla$}{\bf r}_{0} \cdot \mbox{$\boldmath \nabla$}{\bf r}%
_{0}^{\top }$ is employed, where $\top $ stands for transpose. The tensor $%
{\bf A}$ is connected with the strain tensor $\mbox{$\boldmath \epsilon$}$
in the deformed state by the equality $\mbox{$\boldmath \epsilon$}=\frac{1}{2%
}({\bf I}-{\bf A})$, where ${\bf I}$ is the unit tensor. It follows from Eq.
(11) that the non--zero components of $\mbox{$\boldmath \epsilon$}$ are
given by 
\begin{eqnarray}
\epsilon _{13} &=&\epsilon _{31}=-\frac{1}{2} (\xi _{1}\sin \alpha +\xi
_{2}\cos \alpha ) \Bigl(\frac{d\alpha }{d\xi } +\omega -\omega _{0}\Bigr), 
\nonumber \\
\epsilon _{23} &=&\epsilon _{32} =\frac{1}{2} (\xi _{1}\cos \alpha -\xi
_{2}\sin \alpha ) \Bigl(\frac{d\alpha }{d\xi }+\omega -\omega _{0}\Bigr), 
\nonumber \\
\epsilon _{33} &=&-\kappa (\xi _{1}\cos \alpha -\xi _{2}\sin \alpha )
+\kappa_{0}\xi _{1},
\end{eqnarray}
where we kept only terms up to first order in $\xi _{1}$ and $\xi _{2}$. The
neglect of second and higher order terms follows from the standard small
local deformation assumption, which implies that all the length scales
associated with bending, torsion and twist (e.g., radii of curvature) are
much larger than the diameter of the rod. Note that this approximation is
consistent with the form of the displacement field, Eqs. (3) and (7), where
only terms up to linear order in the transverse coordinated $\xi _{1}$ and $%
\xi _{2}$ were kept.

\section{Strain energy density}

For a linear anisotropic elastic medium, the mechanical energy of elongation
per unit volume in the deformed state is calculated as 
\begin{equation}
u_{{\rm el}}=\frac{1}{2}E_{1}\epsilon _{33}^{2},
\end{equation}
and the mechanical energy of shear is 
\begin{equation}
u_{{\rm sh}}=E_{2}(\epsilon _{13}^{2}+\epsilon _{31}^{2}+\epsilon
_{23}^{2}+\epsilon _{32}^{2}),
\end{equation}
where $E_{1}$ and $E_{2}$ are the appropriate elastic moduli. It follows
from Eqs. (12) to (14) that the mechanical energy density 
\begin{equation}
u=u_{{\rm el}}+u_{{\rm sh}}
\end{equation}
is can be written as 
\begin{equation}
u =\frac{1}{2}\biggl \{E_{1}\biggl [\kappa ^{2}(\xi _{1}\cos \alpha -\xi
_{2}\sin \alpha )^{2}+\kappa _{0}^{2}\xi _{1}{}^{2}\Bigr] +E_{2}(\xi
_{1}^{2}+\xi _{2}^{2})\Bigl(\frac{d\alpha }{d\xi }+\omega -\omega
_{0}\Bigr)^{2}\biggr \}.
\end{equation}
The mechanical energy per unit length is given by 
\[
U=\int_{{\cal S}}ud\xi _{1}d\xi _{2}, 
\]
which yields, upon integration 
\begin{equation}
U=\frac{1}{2}\biggl [ A_{1} \Bigl(\kappa ^{2}-2\kappa \kappa _{0}\cos \alpha
+\kappa _{0}^{2}\Bigr)+A_{2}\Bigl(\frac{d\alpha }{d\xi } +\omega -\omega
_{0} \Bigr)^{2}\biggr ]
\end{equation}
with 
\[
A_{1}=E_{1}I, \qquad A_{2}=2E_{2}I, \qquad I=\int_{{\cal S}}\xi
_{1}^{2}d\xi_{1}d\xi _{2} =\int_{{\cal S}}\xi _{2}^{2}d\xi _{1}d\xi _{2},
\qquad \int_{{\cal S}}\xi _{1}\xi _{2}d\xi _{1}d\xi _{2}=0. 
\]
Comparison of Eqs. (1) and (17) shows that the two expressions coincide in
the absence of rotation of the cross--section with respect to the centerline
(no twist, $\alpha =0$). In the general case, when $\alpha \neq 0$, Eq. (17)
differs from Eq. (1) in several important ways:

\begin{enumerate}
\item  The torsion $\omega $ is replaced by $\omega +d\alpha /d\xi $. This
correction has a simple intuitive meaning: the rotation of a point on the
surface of a rod is the sum of the rotation in space of the centerline of
the rod and of the twist of the cross--section about this centerline. Notice
that this correction may always be present, independent of whether the rod
has a non-vanishing spontaneous curvature ($\kappa _{0}$) and spontaneous
torsion ($\omega _{0}$) or not. Such a correction was, in fact, proposed by
previous investigators \cite{GL92}.

\item  The term $2\kappa \kappa _{0}$ is replaced by $2\kappa \kappa
_{0}\cos \alpha $, introducing a non-trivial coupling between the
spontaneous and the actual curvatures of the rod, and the twist of its
cross--section with respect to the centerline. Note that this term appears
only when the rod has a non--vanishing spontaneous curvature and therefore
while it has no effect on the elasticity of straight rods ($\kappa _{0}=0$),
it has a dramatic effect on the elasticity of helices and other curved ($%
\kappa _{0}\neq 0$) rods.

\item  The usual expression for the energy, Eq. (1), is minimized when the
curvature ($\kappa $) and torsion ($\omega $) recover their spontaneous
values ($\kappa _{0}$ and $\omega _{0}$, respectively) in the stress--free
reference state. Although this appears to be no longer true for our energy,
Eq. (17), the difference stems from the fact that we have introduced a new
independent variable ($\alpha $) that describes the twist of the
cross--section with respect to the centerline of the rod. In the absence of
externally applied torques and tensile forces, minimizing the energy with
respect to $\kappa $, $\omega $ and $\alpha $ yields their values in the
stress--free reference state, i.e., $\kappa _{0}$, $\omega _{0}$ and $\alpha
=0$, respectively.
\end{enumerate}

\section{Stress--strain relations}

Denote by ${\mbox{$\boldmath \sigma$}}$ the Cauchy stress tensor and by $%
\sigma ^{ij}$ its contravariant components in the basis of the actual
configuration. Substitution of Eqs. (13) to (15) into the equality 
\[
\sigma ^{ij}=\frac{\partial u}{\partial \epsilon _{ij}} 
\]
results in 
\begin{eqnarray}
\sigma ^{13} &=&\sigma ^{31} =-E_{2}(\xi _{1}\sin \alpha +\xi _{2}\cos
\alpha ) \Bigl(\frac{d\alpha }{d\xi }+\omega -\omega _{0}\Bigr),  \nonumber
\\
\sigma ^{23} &=&\sigma ^{32} =E_{2}(\xi _{1}\cos \alpha -\xi _{2}\cos \alpha
) \Bigl(\frac{d\alpha }{d\xi }+\omega -\omega _{0}\Bigr), \\
\sigma ^{33} &=& E_{1}[-\kappa (\xi _{1}\cos \alpha -\xi _{2}\sin \alpha
)+\kappa _{0}\xi _{1}].
\end{eqnarray}
Equation (19) does not take into account the inextensibility of the
longitudinal axis. In order to enforce this constraint, we add an unknown
parameter $p$ (a Lagrange multiplier analogous to pressure for
incompressible solids) to Eq. (19): 
\begin{equation}
\sigma ^{33}=-p+E_{1}[-\kappa (\xi _{1}\cos \alpha -\xi _{2}\sin \alpha
)+\kappa _{0}\xi _{1}].
\end{equation}
Since the unit normal to a cross-section of the rod coincides with ${\bf t}$%
, the internal force (per unit area) ${\bf f}$ that acts on the
cross--section of the rod is given by 
\begin{equation}
{\bf f}={\bf t}\cdot \mbox{$\boldmath \sigma$}=\sigma ^{13}{\bf n}+\sigma
^{23}{\bf b}+\sigma ^{33}{\bf t}.
\end{equation}
It follows from Eq. (7) that the radius vector $\mbox{$\boldmath \rho$}$
from the center point of the cross--section (its intersection with the
centerline) to an arbitrary point of the cross-section is 
\begin{equation}
\mbox{$\boldmath \rho$}=(\xi _{1}\cos \alpha -\xi _{2}\sin \alpha ){\bf n}%
+(\xi _{1}\sin \alpha +\xi _{2}\cos \alpha ){\bf b}.
\end{equation}
The moment (per unit area) $\mbox{$\boldmath \mu$}$ of the internal force
with respect to the center point\ of the cross--section is defined as 
\begin{equation}
\mbox{$\boldmath \mu$}=\mbox{$\boldmath \rho$}\times {\bf f},
\end{equation}
where $\times $ stands for vector product. Combining Eqs. (21) to (23) and
using the equalities 
\begin{equation}
{\bf t}\times {\bf n}={\bf b},\qquad {\bf n}\times {\bf b}={\bf t},\qquad 
{\bf b}\times {\bf t}={\bf n},
\end{equation}
we find that 
\begin{eqnarray}
\mbox{$\boldmath \mu$} &=&\Bigl[(\xi _{1}\sin \alpha +\xi _{2}\cos \alpha )%
{\bf n}-(\xi _{1}\cos \alpha -\xi _{2}\sin \alpha ){\bf b}\Bigr]\sigma ^{33}
\nonumber \\
&&+\Bigl[(\xi _{1}\cos \alpha -\xi _{2}\sin \alpha )\sigma ^{23}-(\xi
_{1}\sin \alpha +\xi _{2}\cos \alpha )\sigma ^{13}\Bigr]{\bf t}.
\end{eqnarray}
The internal moment ${\bf M}$ is obtained by integrating $%
\mbox{$\boldmath\mu$}$ over the cross--section of the rod, 
\begin{equation}
{\bf M}=\int_{{\cal S}}\mbox{$\boldmath \mu$}d\xi _{1}d\xi _{2}=M_{n}{\bf n}%
+M_{b}{\bf b}+M_{t}{\bf t}.
\end{equation}

In principle, one could proceed in similar fashion and obtain the internal
force 
\begin{equation}
{\bf F=}F_{n}{\bf n}+F_{b}{\bf b}+F_{t}{\bf t}
\end{equation}
by integrating ${\bf f}$ over the cross--section of the rod. However,
inspection of Eqs. (18)--(21) shows that since our expression for ${\bf f}$
is linear in the transverse coordinates $\xi _{1}$ and $\xi _{2}$, the
integral over the cross--section vanishes. The source of the problem can be
traced back to our choice of the displacement fields, Eqs. (3) and (7),
where only linear terms in the transverse coordinates $\xi _{1}$ and $%
\xi_{2} $ were taken into account. Note, however, that even if we were to
keep higher order terms in $\xi _{1}$ and $\xi _{2}$ in these equations, the
unknown function ${\bf F}$ would be expressed in terms of new unknown
functions (coefficients of quadratic contributions in $\xi _{1}$ and $%
\xi_{2} $ to the displacement fields). Instead, we will follow the standard
approach and treat the vector ${\bf F}$ as an additional unknown that is
found from the equilibrium equations (force and moment balance conditions).

We now proceed to calculate the internal moment by substituting expressions
(18) and (21) into Eqs. (25) and (26). Upon integration we obtain the
constitutive relation between the parameters that characterize the
deformation ($\kappa $, $\omega $ and $\alpha $) and the internal moment $%
{\bf M}$ 
\begin{equation}
{\bf M}=A_{1}\kappa _{0}\sin \alpha {\bf n}+A_{1}(\kappa -\kappa _{0}\cos
\alpha ){\bf b}+A_{2}\Bigl(\omega +\frac{d\alpha }{d\xi }-\omega _{0}\Bigr)%
{\bf t}.
\end{equation}
Equation (28) is a new expression for the moment of internal forces which
accounts for the twist of the cross-section with respect to the centerline
of the rod. As expected, the internal moment vanishes in the stress--free
reference state: $\kappa =\kappa _{0}$, $\omega =\omega _{0}$, $\alpha =0$.
In the absence of twist, $\alpha =0$, Eq. (28) reduces to the conventional
expression for Kirchhoff rods 
\begin{equation}
{\bf M}=A_{1}(\kappa -\kappa _{0}){\bf b} +A_{2}\Bigl(\omega -\omega
_{0}\Bigr ){\bf t}.
\end{equation}

\section{Equilibrium equations}

Consider an element of the rod bounded by two cross-sections with
longitudinal coordinates $\xi $ and $\xi +d\xi $. Forces acting on this
element consist of the internal force $-{\bf F}(\xi )$ applied to the
cross-section $\xi $, the internal force ${\bf F}(\xi +d\xi )$ applied to
the cross-section $\xi +d\xi $, and the external force ${\bf q}d\xi $
proportional to the length of the element $d\xi $. Balancing the forces on
the element yields 
\begin{equation}
{\bf F}(\xi +d\xi )-{\bf F}(\xi )+{\bf q}(\xi )d\xi =0.
\end{equation}
Expanding the vector function 
\[
{\bf F}(\xi +d\xi )=F_{n}(\xi +d\xi ){\bf n}(\xi +d\xi )+F_{b}(\xi +d\xi )%
{\bf b}(\xi +d\xi )+F_{t}(\xi +d\xi ){\bf t}(\xi +d\xi ) 
\]
into the Taylor series, using Eq. (5), and neglecting terms of second order
in $d\xi $, we find that 
\begin{equation}
{\bf F}(\xi +d\xi )-{\bf F}(\xi )=\biggl [\Bigl(\frac{dF_{n}}{d\xi }+\kappa
F_{t}-\omega F_{b}\Bigr){\bf n}+\Bigl(\frac{dF_{b}}{d\xi }+\omega F_{n}\Bigr)%
{\bf b}+\Bigl(\frac{dF_{t}}{d\xi }-\kappa F_{n}\Bigr){\bf t}\biggr ]d\xi .
\end{equation}
Substitution of Eq. (31) into Eq. (30) results in the equilibrium equations 
\begin{eqnarray}
&&\frac{dF_{n}}{d\xi }+\kappa F_{t}-\omega F_{b}+q_{n}=0,\qquad \frac{dF_{b}%
}{d\xi }+\omega F_{n}+q_{b}=0, \\
&&\frac{dF_{t}}{d\xi }-\kappa F_{n}+q_{t}=0.
\end{eqnarray}
where $q_{n}$, $q_{b}$, and $q_{t}\ $are the components of the external
force per unit length, ${\bf q}=q_{n}{\bf n}+q_{b}{\bf b}+q_{t}{\bf t}$.

The moments acting on the element of the rod consist of the internal moment $%
-{\bf M}(\xi )$ applied to the cross-section $\xi $, the internal moment $%
{\bf M}(\xi +d\xi )$ applied to the cross-section $\xi +d\xi $, the moments
of internal forces $-{\bf F}(\xi )$ and ${\bf F}(\xi +d\xi )$, and the
external moment ${\bf m}d\xi $ proportional to the length $d\xi $, where $%
{\bf m}=m_{n}{\bf n}+m_{b}{\bf b}+m_{t}{\bf t}$ is the external moment per
unit length. To first order in $d\xi $, the moment of internal forces 
with respect to the center of the cross-section with coordinate $\xi $ is 
\[
\Bigl[{\bf R}(\xi +d\xi )-{\bf R}(\xi )\Bigr]\times {\bf F}(\xi +d\xi )={\bf %
t}(\xi )\times {\bf F}(\xi )d\xi . 
\]
The balance equation for the moments reads 
\begin{equation}
{\bf M}(\xi +d\xi )-{\bf M}(\xi )+{\bf t}(\xi )\times {\bf F}(\xi )d\xi +%
{\bf m}(\xi )d\xi =0.
\end{equation}
It follows from Eqs. (24) and (26) that ${\bf t}\times {\bf F}=-F_{b}{\bf n}%
+F_{n}{\bf b}$. By analogy with Eq. (31), one can write 
\[
{\bf M}(\xi +d\xi )-{\bf M}(\xi )=\biggl [
\Bigl(\frac{dM_{n}}{d\xi }+\kappa M_{t}-\omega M_{b}\Bigr){\bf n}+\Bigl(%
\frac{dM_{b}}{d\xi }+\omega M_{n}\Bigr){\bf b}+\Bigl(\frac{dM_{t}}{d\xi }%
-\kappa M_{n}\Bigr){\bf t}\biggr ]d\xi . 
\]
Substitution of these expressions into Eq. (34) results in the equations 
\begin{eqnarray}
&&\frac{dM_{n}}{d\xi }+\kappa M_{t}-\omega M_{b}-F_{b}+m_{n}=0,\qquad \frac{%
dM_{b}}{d\xi }+\omega M_{n}+F_{n}+m_{b}=0, \\
&&\frac{dM_{t}}{d\xi }-\kappa M_{n}+m_{t}=0.
\end{eqnarray}
Given the vectors ${\bf M}$ and ${\bf m}$, Eqs. (35) can be used to
determine the forces $F_{n}$ and $F_{b}$. Eliminating the unknown functions $%
F_{n}$ and $F_{b}$ from Eqs. (32) and (35), we obtain 
\begin{eqnarray}
&&\frac{dF_{t}}{d\xi }+\kappa \biggl (\frac{dM_{b}}{d\xi }+\omega M_{n}+m_{b}%
\biggr )+q_{t}=0, \\
&&\frac{d}{d\xi }\biggl (\frac{dM_{b}}{d\xi }+\omega M_{n}+m_{b}\biggr )%
+\omega \biggl (\frac{dM_{n}}{d\xi }+\kappa M_{t}-\omega M_{b}+m_{n}\biggr )%
-\kappa F_{t}-q_{n}=0,  \nonumber \\
&&\frac{d}{d\xi }\biggl (\frac{dM_{n}}{d\xi }+\kappa M_{t}-\omega M_{b}+m_{n}%
\biggr )-\omega \biggl (\frac{dM_{b}}{d\xi }+\omega M_{n}+m_{b}\biggr )%
+q_{b}=0.
\end{eqnarray}
Equations (36) to (38) together with constitutive relation (28) are a set of
four nonlinear differential equations which determine the four unknown
functions $F_{t}$, $\alpha $, $\kappa $ and $\omega $. The neglect of $%
\alpha $ (that is the use of conventional formula (6) instead of Eq. (7) for
the displacement field ${\bf r}$) is acceptable only when special
restrictions are imposed on external forces and moments. In the general
case, this simplification is not correct, and Eq. (7) should be employed for
the analysis of deformations.

\section{Examples}

\subsection{Twist of a closed loop}

Consider a rod whose stress-free shape is a planar circular loop with radius 
$a_{0}$, under the action of a constant twisting moment $m_{t}$. It is
assumed that the moments $m_{n}$ and $m_{b}$, as well as the forces $q_{n}$, 
$q_{b}$ and $q_{t}$ vanish. The solution of Eqs. (36) to (38) reads 
\begin{eqnarray}
&&\kappa =\kappa _{0}=a_{0}^{-1}, \\
&&\omega =\omega _{0}=0,\qquad F_{t}=0,\qquad \alpha =\arcsin \frac{%
m_{t}a_{0}^{2}}{A_{1}}.
\end{eqnarray}
According to these equalities, any cross-section of the rod twists around
its centerline by a constant angle $\alpha $. This solution is not described
by the Kirchhoff theory of thin rods. It exists as long as the moment $m_{t}$
satisfies the condition $|m_{t}|\leq A_{1}a_{0}^{-2}$. If the latter
restriction is not fulfilled, the planar shape of the loop becomes unstable.

\subsection{Torsion of a disconnected ring}

We analyze the deformation of a disconnected ring (no contact between the
points $\xi =0$ and $\xi =L$). The end $\xi =0$ is fixed, and a torque $T$
is applied to the free end $\xi =L$. The centerline of the rod in the
stress--free reference state describes a planar circle with radius $%
a_{0}=\kappa _{0}^{-1}$ and no spontaneous torsion, $\omega _{0}=0$. Similar
problems were recently studied in \cite{GL92,GT97,HZ99,WT86} and their
solutions were applied to the analysis of kink transitions in short DNA
rings.

We assume that in the deformed state the centerline becomes a non-planar
curve whose radius of curvature remains unchanged, see Eq. (39). For
simplicity, we confine ourselves to small displacements, and neglect terms
of order $\alpha ^{2}$ in the constitutive equations (28). This yields 
\begin{equation}
M_{n}=A_{1}\kappa _{0}\alpha ,\qquad M_{b}=0,\qquad M_{t}=A_{2}\Bigl(\frac{%
d\alpha }{d\xi }+\omega \Bigr).
\end{equation}
Substitution of these expressions into the equilibrium equations (36) to
(38) implies that the longitudinal force $F_{t}$ vanishes, whereas the
functions $\alpha $ and $\omega $ obey the equations 
\begin{equation}
A_{2}\Bigl(\frac{d^{2}\alpha }{d\xi ^{2}}+\frac{d\omega }{d\xi }%
\Bigr)-A_{1}\kappa _{0}^{2}\alpha =0,\qquad \frac{d}{d\xi }\biggl [A_{1}%
\frac{d\alpha }{d\xi }+A_{2}\Bigl(\frac{d\alpha }{d\xi }+\omega \Bigr)\biggr
]=0.
\end{equation}
It follows from the second equality in Eq. (42) that 
\begin{equation}
(A_{1}+A_{2})\frac{d\alpha }{d\xi }+A_{2}\omega =c,
\end{equation}
where $c$ is a constant to be found. Excluding $\omega $ from Eqs. (42) and
(43), we obtain 
\begin{equation}
\frac{d^{2}\alpha }{d\xi ^{2}}+\kappa _{0}^{2}\alpha =0.
\end{equation}
The solution of Eq. (44) is given by 
\begin{equation}
\alpha =c_{1}\sin \kappa _{0}\xi +c_{2}\cos \kappa _{0}\xi ,
\end{equation}
where $c_{1}$ and $c_{2}$ are arbitrary constants. Substitution of Eqs. (43)
and (45) into the boundary conditions at the clamped end $\xi =0$ 
\[
\alpha (0)=0,\qquad \omega (0)=0 
\]
implies that 
\begin{equation}
\alpha =c_{1}\sin \kappa _{0}\xi ,\qquad \omega =\frac{A_{1}+A_{2}}{A_{2}}%
\kappa _{0}c_{1}(1-\cos \kappa _{0}\xi ).
\end{equation}
Equating the moment $M_{t}$ at the end $\xi =L$ to the external torque $T$
and using Eqs. (41) and (46), we obtain 
\[
c_{1}=\frac{T}{A_{2}\kappa _{0}}, 
\]
which results in the formulas 
\begin{equation}
\alpha =\frac{T}{A_{2}\kappa _{0}}\sin (\kappa _{0}\xi ),\qquad \omega =%
\frac{T(A_{1}+A_{2})}{A_{2}^{2}}\Bigl(1-\cos (\kappa _{0}\xi )\Bigr).
\end{equation}
Equations (41) and (47) provide an explicit solution to the torque problem,
which cannot be obtained in the framework of the Kirchhoff theory of rods.
When the radius of the ring tends to infinity, i.e. for a prismatic rod, Eq.
(47) implies that 
\begin{equation}
\alpha =\frac{T}{A_{2}}\xi ,\qquad \omega =0.
\end{equation}
In this limit, the solution (48) coincides with the classical displacement
field for the twist of a circular cylinder \cite{Dro96}.

\subsection{Helix under tension and torque}

A helix--shaped rod whose stress--free reference state is characterized by
spontaneous curvature $\kappa _{0}$ and torsion $\omega _{0}$, is deformed
by tensile forces $P$ and torques $T$ applied to its ends. All other forces $%
{\bf q}$ and moments ${\bf m}$ are assumed to vanish. We introduce Cartesian
coordinates $\{x_{k}\}$ with unit vectors ${\bf e}_{k}$ and describe the
configuration of the centerline of the rod in the stress--free reference
state by the vector 
\begin{equation}
{\bf R}_{0}=a_{0}\cos \frac{\xi }{\sqrt{a_{0}^{2}+b_{0}^{2}}}{\bf e}%
_{1}+a_{0}\sin \frac{\xi }{\sqrt{a_{0}^{2}+b_{0}^{2}}}{\bf e}_{2}+\frac{%
b_{0}\xi }{\sqrt{a_{0}^{2}+b_{0}^{2}}}{\bf e}_{3}.
\end{equation}
The parameters $a_{0}$ and $b_{0}$ are expressed in terms of the spontaneous
curvature $\kappa _{0}$ and torsion $\omega _{0}$ by the formulas 
\begin{equation}
\kappa _{0}=\frac{a_{0}}{a_{0}^{2}+b_{0}^{2}},\qquad \omega _{0}=\frac{b_{0}%
}{a_{0}^{2}+b_{0}^{2}}.
\end{equation}

\subsubsection{Fixed force and torque on ends}

Consider a rod whose centerline describes one complete turn of a helix (the
angle between tangent vectors at the two ends of the undeformed rod equals $%
2\pi ).$ The contour length of the rod is 
\begin{equation}
l=2\pi (\kappa _{0}^{2}+\omega _{0}^{2})^{-\frac{1}{2}}.
\end{equation}
We assume the following boundary conditions at the ends of the rod: 
\begin{eqnarray}
&&M_{n}(0)=M_{n}(l)=0,\qquad M_{b}(0)=M_{b}(l)=0,  \nonumber \\
&&M_{t}(0)=M_{t}(l)=T,\qquad F_{t}(0)=F_{t}(l)=P.
\end{eqnarray}
Equations (52) imply that the torque $T$ and the tensile force $P$ are the
only external loads applied to the segment. Assuming the parameters $P$ and $%
T$ to be rather small and neglecting the deviation of torsion from its value
in the stress--free state, we look for a solution of the equilibrium
equations in the form 
\begin{equation}
\alpha =\Delta \alpha ,\qquad \kappa =\kappa _{0}+\Delta \kappa ,\qquad
\omega =\omega _{0},
\end{equation}
where $\Delta \alpha $ is small compared to unity, and $\Delta \kappa $ is
small compared to $\kappa _{0}$.

Neglecting terms of the second order in the perturbations of twist angle and
curvature ($\Delta \alpha \ $and $\Delta \kappa $, respectively), we find
from Eq. (28) that 
\begin{equation}
M_{n}=A_{1}\kappa _{0}\Delta \alpha ,\qquad M_{b}=A_{1}\Delta \kappa ,\qquad
M_{t}=A_{2}\frac{d\Delta \alpha }{d\xi }.
\end{equation}
We substitute expressions (53) and (54) into Eqs. (36) to (38), neglect
terms of the second order in $\Delta \alpha \ $and $\Delta \kappa $, and
arrive at the equations 
\begin{eqnarray}
&&\frac{dM_{t}}{d\xi }-\kappa _{0}M_{n}=0, \\
&&\frac{dF_{t}}{d\xi }+\kappa _{0}\Bigl(\frac{dM_{b}}{d\xi }+\omega
_{0}M_{n}\Bigr)=0, \\
&&\frac{d^{2}M_{b}}{d\xi ^{2}}+2\omega _{0}\frac{dM_{n}}{d\xi }+\kappa
_{0}\omega _{0}M_{t}-\omega _{0}^{2}M_{b}-\kappa _{0}F_{t}=0, \\
&&\frac{d^{2}M_{n}}{d\xi ^{2}}+\kappa _{0}\frac{dM_{t}}{d\xi }-2\omega _{0}%
\frac{dM_{b}}{d\xi }-\omega _{0}^{2}M_{n}=0,
\end{eqnarray}
where the longitudinal force $F_{t}$ is assumed to be small as well. It
follows from Eqs. (55) and (58) that 
\begin{equation}
\frac{dM_{b}}{d\xi }=\frac{1}{2\omega _{0}}\biggl [\frac{d^{2}M_{n}}{d\xi
^{2}}+(\kappa _{0}^{2}-\omega _{0}^{2})M_{n}\biggr ].
\end{equation}
Substitution of Eq. (59) into Eq. (56) results in 
\begin{equation}
\frac{dF_{t}}{d\xi }+\frac{\kappa _{0}}{2\omega _{0}}\biggl [\frac{d^{2}M_{n}%
}{d\xi ^{2}}+(\kappa _{0}^{2}+\omega _{0}^{2})M_{n}\biggr ]=0.
\end{equation}
Equations (55), (57) and (59) imply that 
\begin{eqnarray}
\frac{dF_{t}}{d\xi } &=&\frac{1}{\kappa _{0}}\biggl (\frac{d^{3}M_{b}}{d\xi
^{3}}+2\omega _{0}\frac{d^{2}M_{n}}{d\xi ^{2}}+\kappa _{0}\omega _{0}\frac{%
dM_{t}}{d\xi }-\omega _{0}^{2}\frac{dM_{b}}{d\xi }\biggr )  \nonumber \\
&=&\frac{1}{2\kappa _{0}\omega _{0}}\biggl [\frac{d^{4}M_{n}}{d\xi ^{4}}%
+(\kappa _{0}^{2}+2\omega _{0}^{2})\frac{d^{2}M_{n}}{d\xi ^{2}}+\omega
_{0}^{2}(\kappa _{0}^{2}+\omega _{0}^{2})M_{n}\biggr ].
\end{eqnarray}
Excluding the function $F_{t}$ from Eqs. (60) and (61), we obtain a closed
equation for the internal moment $M_{n}$ 
\begin{equation}
\frac{d^{4}M_{n}}{d\xi ^{4}}+2(\kappa _{0}^{2}+\omega _{0}^{2})\frac{%
d^{2}M_{n}}{d\xi ^{2}}+(\kappa _{0}^{2}+\omega _{0}^{2})^{2}M_{n}=0.
\end{equation}
The solution of Eq. (62) reads 
\begin{equation}
M_{n}=(c_{1}+c_{1}^{\prime }\xi )\sin \Bigl(\sqrt{\kappa _{0}^{2}+\omega
_{0}^{2}}\xi \Bigr)+(c_{2}+c_{2}^{\prime }\xi )\cos \Bigl(\sqrt{\kappa
_{0}^{2}+\omega _{0}^{2}}\xi \Bigr),
\end{equation}
where $c_{k}$, $c_{k}^{\prime }$ are constants to be found. It follows from
the boundary conditions (52) for the function $M_{n}$ and Eq. (63) that 
\begin{equation}
c_{2}=c_{2}^{\prime }=0.
\end{equation}
Integrating Eq. (59) from 0 to $l$ and using boundary conditions (52) for
the function $M_{b}$, we obtain 
\[
\int_{0}^{l}\biggl [\frac{d^{2}M_{n}}{d\xi ^{2}}+(\kappa _{0}^{2}-\omega
_{0}^{2})M_{n}\biggr ]d\xi =0. 
\]
Substitution of expressions (63) and (64) into this equality results in 
\begin{equation}
c_{1}^{\prime }=0.
\end{equation}
Combining Eqs. (54) and (63) to (65), we find that 
\begin{equation}
\Delta \alpha (\xi )=\frac{c_{1}}{A_{1}\kappa _{0}}\sin \Bigl(\sqrt{\kappa
_{0}^{2}+\omega _{0}^{2}}\xi \Bigr),
\end{equation}
Note that although the twist angle vanishes at the ends and in the middle of
the rod ($\Delta \alpha (0)=\Delta \alpha (l)=\Delta \alpha (l/2)=0$), it
does not vanish elsewhere. Differentiating Eq. (66) and using Eq. (54) and
the boundary conditions (52) for $M_{t}$, we arrive at the equality 
\begin{equation}
c_{1}=\frac{A_{1}T\kappa _{0}}{A_{2}\sqrt{\kappa _{0}^{2}+\omega _{0}^{2}}}.
\end{equation}
Substitution of Eqs. (64), (65) and (67) into Eqs. (54), (63) and (66)
implies that 
\begin{equation}
M_{n}=\frac{A_{1}T\kappa _{0}}{A_{2}\sqrt{\kappa _{0}^{2}+\omega _{0}^{2}}}%
\sin \Bigl(\sqrt{\kappa _{0}^{2}+\omega _{0}^{2}}\xi \Bigr),\qquad
M_{t}=T\cos \Bigl(\sqrt{\kappa _{0}^{2}+\omega _{0}^{2}}\xi \Bigr).
\end{equation}
It follows from Eqs. (59) and (68) that 
\[
\frac{dM_{b}}{d\xi }=-\frac{A_{1}T\kappa _{0}\omega _{0}}{A_{2}\sqrt{\kappa
_{0}^{2}+\omega _{0}^{2}}}\sin \Bigl(\sqrt{\kappa _{0}^{2}+\omega _{0}^{2}}%
\xi \Bigr). 
\]
Integrating this equality with the boundary conditions (52) and substituting
in Eq. (54) yields 
\begin{equation}
\Delta \kappa (\xi )=\frac{M_{b}}{A_{1}}=-\frac{T\kappa _{0}\omega _{0}}{%
A_{2}(\kappa _{0}^{2}+\omega _{0}^{2})}\biggl [1-\cos \Bigl(\sqrt{\kappa
_{0}^{2}+\omega _{0}^{2}}\xi \Bigr)\biggr ]
\end{equation}
Note that the sign of $\Delta \kappa $ vanishes at the ends of the rod;
inside it, its sign is opposite to that of the torque $T$ (positive torque
means overtwisting). Substitution of Eqs. (68) and (69) into Eq. (57) gives
the internal tensile force 
\begin{equation}
F_{t}=T\omega _{0}\biggl [\biggl (1+\frac{A_{1}}{A_{2}}\Bigl(1+\frac{\omega
_{0}^{2}}{\kappa _{0}^{2}+\omega _{0}^{2}}\Bigr)\biggr )\cos \Bigl(\sqrt{%
\kappa _{0}^{2}+\omega _{0}^{2}}\xi \Bigr)-\frac{A_{1}\omega _{0}^{2}}{%
A_{2}(\kappa _{0}^{2}+\omega _{0}^{2})}\biggr ].
\end{equation}
It follows from Eq. (70) that our solution $\kappa (\xi )$, $\omega $ and $%
\alpha (\xi )$ under boundary conditions (52) is valid if the tensile force $%
P$ and the torque $T$ applied to the ends of the rod satisfy the relation 
\begin{equation}
P=T\omega _{0}\Bigl(1+\frac{A_{1}}{A_{2}}\Bigr).
\end{equation}
Equations (68) to (70) provide an explicit solution to the problem of
combined tension and torsion of a helical segment. The main results are as
follows:

\begin{enumerate}
\item  The application of positive torque $T$ at the ends (overtwist) leads
to axial compression of the helix which is maximal at the center and
vanishes at the ends of the rod;

\item  The ratio of the tensile force $P$ and the torque $T$ is independent
of the initial curvature $\kappa _{0}$ (and, therefore, of the length of the
rod) and depends only on the initial torsion $\omega _{0}$ and the ratio of
elastic moduli $A_{1}/A_{2}=E_{1}/(2E_{2})$;

\item  The force $P$ is proportional to the torque $T.$ This result is
markedly different from that obtained for a similar deformation of a
circular incompressible cylinder, where $P$ can be shown to be proportional
to $T^{2}$ (the Poynting effect\cite{Dro96}).
\end{enumerate}

Note that our solution corresponds to a helical rod (with constant $\kappa
_{0}$ and $\omega _{0}$) which, upon application of external forces and
torques, is deformed into a new, non-helical shape. It is natural to ask
under which boundary conditions a helix will deform into another helix (with
constant $\kappa $ and $\omega $), and derive the corresponding
force--elongation relation. This is done in the following.

\subsubsection{Elongation and winding of a helix}

Consider a helix made of an arbitrary number of repetitive units ${\cal L}%
_{0}$ (${\cal L}_{0}$ is the smallest segment for which the angle between
tangent vectors at its ends is $2\pi $) such that its stress--free reference
state is characterized by the parameters $\kappa _{0}$ and $\omega _{0}$. We
allow deformations that satisfy the following conditions: (i) the rod
becomes a helix with constant curvature $\kappa $ and constant torsion $%
\omega $, and (ii) the twist $\alpha $ vanishes. Under the action of
combined tensile force $P$ and torque $T$, any repetitive unit ${\cal L}_{0}$
of the helix in the reference state is transformed into an element with the
angle between tangent vectors at the ends $2\pi (1+\varphi )$, where the
angle $2\pi \varphi $ can take positive or negative values. The radius
vector of the centerline of the rod in the deformed state can be written in
the form 
\begin{equation}
{\bf R}=a\cos (S\xi ){\bf e}_{1}+a\sin (S\xi ){\bf e}_{2}+S_{1}\xi {\bf e}%
_{3},
\end{equation}
where $a$, $S$ and $S_{1}$ are constants which will be calculated in the
following. Differentiating Eq. (72) with respect to $\xi $ and bearing in
mind that $|{\bf t}|=1$, we obtain 
\begin{equation}
a^{2}S^{2}+S_{1}^{2}=1.
\end{equation}
According to the definition of $\varphi $, 
\begin{equation}
Sl=2\pi (1+\varphi ).
\end{equation}
The projected distances (along the $x_{3}$-axis) between the ends of the
repetitive unit in the reference and deformed states are 
\begin{equation}
\Pi _{0}=2\pi b_{0},\qquad \Pi =S_{1}l,
\end{equation}
respectively. The axial elongation $\eta $ is defined as the ratio of these
distances, 
\begin{equation}
\eta =\frac{\Pi }{\Pi _{0}}=\frac{S_{1}l}{2\pi b_{0}}=\frac{l}{2\pi b_{0}}%
\sqrt{1-a^{2}S^{2}}.
\end{equation}
Simple calculations result in the formulas 
\begin{equation}
\kappa =aS^{2},\qquad \omega =S\sqrt{1-a^{2}S^{2}}.
\end{equation}
It follows from Eq. (72) that the projection of the force $F_{t}$ on the
axis $x_{3}$ is $F_{t}{\bf t}\cdot {\bf e}_{3}=F_{t}S_{1}$. Equating this
expression to the tensile force $P$ and using Eq. (73), we arrive at the
relation 
\begin{equation}
F_{t}=\frac{P}{\sqrt{1-a^{2}S^{2}}},
\end{equation}
which means that $F_{t}$ is independent of $\xi $. Equation (28) implies
that components of the moment ${\bf M}$ are independent of $\xi $ as well, 
\begin{equation}
M_{n}=0,\qquad M_{b}=A_{1}(\kappa -\kappa _{0}),\qquad M_{t}=A_{2}(\omega
-\omega _{0}).
\end{equation}
The only equilibrium equation reads 
\[
\omega (\kappa M_{t}-\omega M_{b})=\kappa F_{t}. 
\]
Substitution of expressions (78) and (79) into this equality yields 
\begin{equation}
\omega \Bigl[A_{2}\kappa (\omega -\omega _{0})-A_{1}\omega (\kappa -\kappa
_{0})\Bigr]=\frac{\kappa P}{\sqrt{1-a^{2}S^{2}}}.
\end{equation}
Excluding the parameters $a$, $S$, $\kappa $ and $\omega $ from Eqs. (73),
(74), (76), (77) and (80), we express the tensile force $P$ in terms of the
axial elongation of the helix $\eta $: 
\begin{equation}
P_{0}=\frac{\lambda (1+\varphi )\eta ^{2}}{\sqrt{1+\lambda ^{2}}}\biggl [%
(1+\varphi )\eta -1-A\eta \Bigl((1+\varphi )-\frac{1}{\sqrt{1+\lambda
^{2}(1-\eta ^{2})}}\Bigr)\biggr ],
\end{equation}
where 
\[
A=\frac{A_{1}}{A_{2}},\qquad \lambda =\frac{\omega _{0}}{\kappa _{0}},\qquad
P_{0}=\frac{P}{A_{2}\omega _{0}^{2}}. 
\]

Comparing Eq. (52) with Eqs. (78) and (79) we find that the only difference
between the boundary conditions for the two problems is that in the former
case we have neglected the moment $M_{b}$. The fact that a minor change of
boundary conditions can drastically change the character of deformation is
quite remarkable and indicates that these conditions should be chosen with
care.

Since various variants of the theory of elastic rods were applied to
interpret the experimental force--elongation curves for stretched DNA
molecules at large deformations, we will present plots of some of the
results of this section and comment on their qualitative features. The graph 
$P_{0}=P_{0}(\eta )$ for extension without torsion, $\varphi =0$, is plotted
in Figure~1. In the calculation we used $A=0.67$, in agreement with
conventional data on DNA $\tilde{A}_{1}=50$ nm, $\tilde{A}_{2}=75$ nm \cite
{Mar98}, where $\tilde{A}_{k}=A_{k}/(k_{B}T)$, $k_{B}$ is Boltzmann's
constant and $T$ is temperature. No detailed comparison with experiment is
attempted here, but Figure~1 captures rather well the qualitative features
of the experimental data for DNA molecules \cite{SFB92,SCB96}.

In order to check whether our theory captures the qualitative features of
experimental data on the elasticity of supercoiled DNA, the dependence $\eta
=\eta (\varphi )$ is depicted in Figure~2 for various tensile forces $P_{0}$%
. This figure also shows qualitative agreement with observations on the DNA
chains \cite{MN98,SAB96}: for small tensile forces, there is pronounced
asymmetry with regard to the sign of $\varphi$, but the $\eta =\eta (\varphi
)$ curve becomes nearly flat at large tensile forces. Throughout the
parameter range, the elongation decreases nearly linearly with degree of
supercoiling. All these features were observed experimentally and were
interpreted as a proof for the existence of a new type of twist--stretch
coupling\cite{KLN97,Mar97}. Note, however, that in the analysis that led to
Figure 2 we assumed that the deformation of the helix takes place with no
twist of its cross--section around the centerline of the rod ($\alpha =0$).
Therefore, our solution can be derived using the standard theory of elastic
rods, based on the elastic energy of Eq. (1), in which no such coupling
appears. Inspection of the derivation of Eq. (81) leads to the conclusion
that the strong dependence of elongation on the degree of supercoiling has a
simple physical meaning: when an inextensible helical rod is subjected to
torque that produces supercoiling, each new turn has non-vanishing
projection on the $x_{1}-x_{2}$ plane and the projection of the deformed
helix on the $x_{3}$ axis (i.e., its elongation) decreases as the result.
This can be fully described by Eq. (1) and does not require the introduction
of new coupling into the mechanical energy of elastic rods.

\section{Concluding remarks}

In this work we have extended the theory of elasticity of thin inextensible
rods beyond that of three--dimensional space curves which can be completely
described by local curvature $\kappa $ and geometric torsion $\omega $. We
have shown that in order to describe the displacement of a point in a rod of
arbitrarily small but non--vanishing thickness, one has to account for
deformations that produce a rotation of the cross--section of the rod about
its centerline. The modified displacement field was then used to calculate
the strain tensor. The resulting expression for the mechanical energy of
rods with non--vanishing spontaneous curvature contains a new coupling term
between the curvature of the rod and the twist of its cross--section with
respect to the centerline, which does not appear in any of the previous
theories. We derived the complete set of non-linear differential equations
which describe the conditions of mechanical equilibrium and which can be
solved for the parameters of deformation $\kappa $, $\omega $ and $\alpha $
for arbitrary external forces and moments acting on the rod. In order to
illustrate the physical consequences of our theory, we proceeded to analyze
several illustrative examples. In particular, we have analyzed the
deformation of a helical rod subjected to a combination of tension and
torque and showed that the theory captures the qualitative features of the
recent observations on the connection between supercoiling and elongation of
strongly stretched DNA molecules.

Note that we have described the deformation of thin rods by three
independent functions $\alpha $, $\kappa $ and $\omega $. This is
reminiscent of the conventional approach\cite{GT97} where the deformation is
described in terms of the three components of the, so called, ``twist''
vector, $\kappa _{1},$ $\kappa _{2}$ and $\kappa _{3}$. Although this was
not mentioned by the above authors, such an approach goes beyond the purely
geometric description of an elastic line in which only two functions are
necessary\cite{SH94} \ and describes a line with some ``internal
structure''. With each point of this line one can associate a ``physical''
triad of vectors that differs, in general, from the ``geometric'' (Frenet)
triad. While the two triads have one common vector (the tangent to the
line), the other two pairs of vectors rotate at different rates as one moves
along the line contour and therefore the rotation of the physical triad can
not be completely described by the two Frenet parameters $\kappa $ and $%
\omega $. It is important to realize that the introduction of a physical
triad is necessary whenever some asymmetry of the cross--section, either
geometric or physical\cite{LGK98}, is present. However, even though the
procedure is not unique, one may also introduce the physical triad by hand
even for a rod with a circular cross--section. For example, we may draw a
line on the surface of a rod which describes the intersection of the normal
vector with this surface. When the rod is deformed, the deformation of this
line will, in general, be different from that of the centerline. We can now
connect the corresponding points of the two lines (having the same contour
parameter $\xi $) \ and define the resulting vector as one of the vectors of
the physical triad. The remaining vector is then defined as the normal to
the plane formed by the above vector and the tangent to the centerline. This
procedure is completely equivalent to what we have done here, by introducing
the rotation $\alpha (\xi )$ and explains the appearance of an $\alpha -$%
dependent term ($\kappa $ $\kappa _{0}\cos \alpha $) in the expression for
the mechanical energy, that couples the curvatures in the stress--free and
the deformed states of the rod. Note that while for rods with asymmetric
cross--sections, three independent parameters are needed in order to
characterize the stress--free reference state, only two such parameters
(e.g., $\kappa _{0}$ and $\omega _{0}$) are necessary in the degenerate case
of rods with circular cross--sections.

There are several possible directions in which the work presented here can
be extended. For example, throughout this work we assumed that the
conditions of mechanical equilibrium can be satisfied and considered only
stable configurations of the deformed rods. However, the introduction of a
new type of deformations is expected to have a profound effect on various
instabilities (e.g., buckling under torsion and twist, plectoneme formation,
etc.) and we are now studying these questions. Another direction for future
research involves the extension of the present, purely mechanical, analysis
to include the effects of thermal fluctuations. This leads naturally to a
new class of physical models for rigid biopolymers and protein assemblies
which can account for the spontaneous curvature\cite{TTH87} of these objects.

\acknowledgments
We would like to thank M. Elbaum and D. Kessler for helpful discussions and
suggestions. AD gratefully acknowledges financial support by the Israeli
Ministry of Science through grant 1202--1--98. YR would like to acknowledge
financial support by a grant from the Israel Science Foundation.

\newpage \baselineskip = .5\baselineskip  

\begin{figure}[tbp]
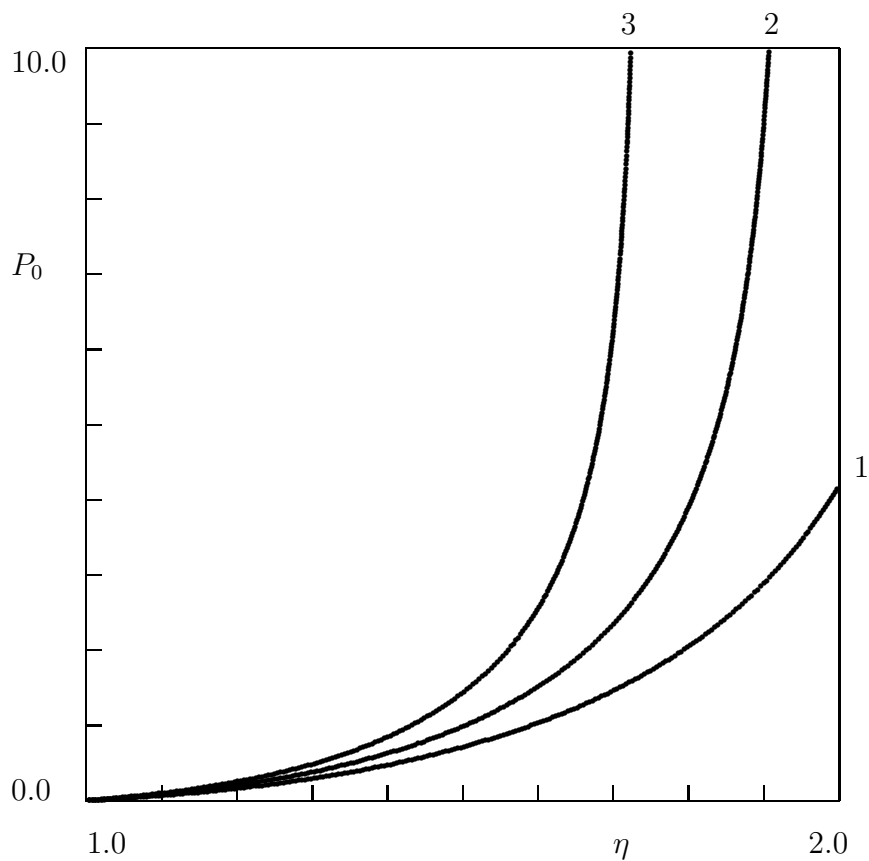

\caption{The dimensionless tensile force $P_{0}$ versus the axial elongation 
$\protect\eta $ for a helix with $A=0.67$ and $\protect\varphi =0$. Curve~1: 
$\protect\lambda =0.5$; curve~2: $\protect\lambda =0.6$; curve~3: $\protect%
\lambda =0.7$}
\end{figure}

\begin{figure}[tbp]
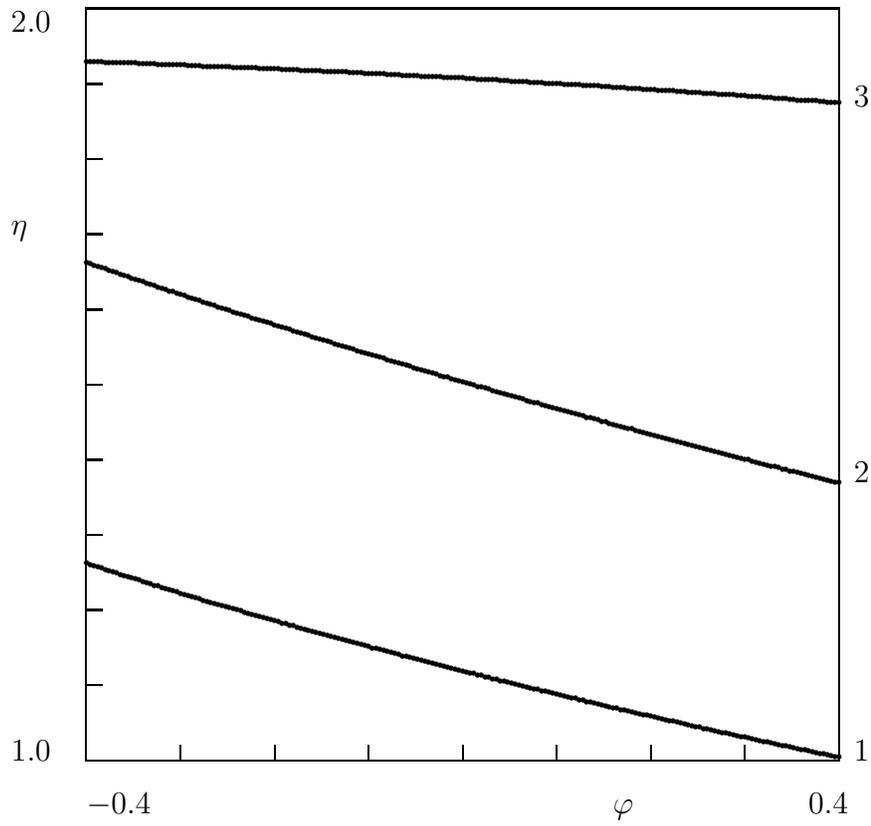

\caption{The axial elongation $\protect\eta$ versus the overtwist $\protect%
\varphi$ for a helix with $A=0.67$ and $\protect\lambda=0.6$. Curve~1: $%
P_{0}=0.1$; curve~2: $P_{0}=1.0$; curve~3: $P_{0}=10.0$}
\end{figure}
\newpage

\setcounter{figure}{0}
\setlength{\unitlength}{1.0 mm}
\begin{figure}[t]
\begin{center}

\end{center}
\vspace*{20 mm}

\caption{The dimensionless tensile force $P_{0}$
versus the axial elongation $\eta$ for a helix
with $A=0.67$ and $\varphi=0$.
Curve~1: $\lambda=0.5$;
curve~2: $\lambda=0.6$;
curve~3: $\lambda=0.7$}
\end{figure}

\begin{figure}[t]
\begin{center}
%
\end{center}
\vspace*{20 mm}

\caption{The axial elongation $\eta$ versus the
overtwist $\varphi$ for a helix with $A=0.67$ and $\lambda=0.6$.
Curve~1: $P_{0}=0.1$;
curve~2: $P_{0}=1.0$;
curve~3: $P_{0}=10.0$}
\end{figure}

\end{document}